\pdfoutput=1

\documentclass[a4paper, 10pt, conference]{IEEEtran} 
\usepackage{times}
\usepackage{graphicx}
\usepackage{epstopdf}
\usepackage{multirow, makecell}
\usepackage{balance}
\usepackage{enumitem}
\usepackage[bookmarks=false]{hyperref}
\usepackage{url}

\usepackage{amsmath}

\newcommand\blfootnote[1]{%
  \begingroup
  \renewcommand\thefootnote{}\footnote{#1}%
  \addtocounter{footnote}{-1}%
  \endgroup
}

\renewcommand{\footnoterule}{%
  \kern -3pt
  \hrule width \columnwidth height 1pt
  \kern 2pt
}

\usepackage{color}
\begin{document}

\title{\textbf{A Reproducible Analysis of RSSI Fingerprinting for Outdoor Localization Using Sigfox: Preprocessing and Hyperparameter Tuning   }\\}

\author{\IEEEauthorblockN{~\large Grigorios G. Anagnostopoulos, Alexandros Kalousis}
\IEEEauthorblockA{Geneva School of Business Administration, HES-SO
Geneva, Switzerland\\
Email:  $\left\{ grigorios.anagnostopoulos, alexandros.kalousis \right\}$@hesge.ch}
}

\maketitle

\begin{abstract}
Fingerprinting techniques, which are a common method for indoor localization, have been recently applied with success into outdoor settings. Particularly, the communication signals of Low Power Wide Area Networks (LPWAN) such as Sigfox, have been used for localization. In this rather recent field of study, not many publicly available datasets, which would facilitate the consistent comparison of different positioning systems, exist so far. 
 In the current study, a published dataset of RSSI measurements on a Sigfox network deployed in Antwerp, Belgium is used to analyse the appropriate selection of preprocessing steps and to tune the hyperparameters of a kNN fingerprinting method. Initially, the tuning of hyperparameter $k$ for a variety of distance metrics, and the selection of efficient data transformation schemes, proposed by relevant works, is presented. In addition, accuracy improvements are achieved in this study, by a detailed examination of the appropriate adjustment of the parameters of the data transformation schemes tested, and of the handling of out of range values.  With the appropriate tuning of these factors, the achieved mean localization error was 298 meters, and the median error was 109 meters. To facilitate the reproducibility of tests and comparability of results, the code and train/validation/test split used in this study are available.
\end{abstract}

\renewcommand\IEEEkeywordsname{Keywords}

\IEEEpeerreviewmaketitle

\blfootnote{\\
Code: \url{https://doi.org/10.5281/zenodo.3228752}\\
Data: \url{https://doi.org/10.5281/zenodo.3228744}\\\\
Preprint\\
\copyright 2019 IEEE.  Personal use of this material is permitted.  Permission from IEEE must be obtained for all other uses, in any current or future media, including reprinting/republishing this material for advertising or promotional purposes, creating new collective works, for resale or redistribution to servers or lists, or reuse of any copyrighted component of this work in other works.
}

%
%	  \IEEEoverridecommandlockouts
%	  \IEEEpubid{\makebox[\columnwidth]
%	  {\hfill  978-1-7281-1788-1/19/\$31.00~\copyright~2019 IEEE}
%	  \hspace{\columnsep}\makebox[\columnwidth]{ }}

%xxx-x-xxxx-xxxx-x-xx-/\$31.00 \textcopyright 2019 IEEE}
% main text

\section{Introduction} \label{sec:Introduction}

%\textcolor{blue}{24/5/2019 1 page}

The recent emergence of Internet of Things (IoT) technologies has made so that a plethora of low power devices make their appearance worldwide, in people's everyday life. The concept of smart cities becomes familiar to the broad public, and numerous applications are being proposed, implemented and deployed in domains such as massive gathering of sensor measurements, automatic control, asset tracking, etc. One domain of great interest concerning the IoT technologies is the offering of Location Based Services.

Outdoor positioning is generally considered a solved problem, as various Global Navigation Satellite Systems (GNSS), such as the commonly known Global Positioning System (GPS), Galileo, GLONASS and BeiDou, have made outdoors positioning an everyday reality for most users of smartphones and custom positioning devices. These systems achieve an impressive accuracy in their estimates. Nevertheless, the battery consumption of their chipsets is considerable, and when it comes to low power IoT mobile devices, their usage is problematic. Therefore, an alternative way of localizing such devices is needed.

The proliferation of IoT devices has been facilitated by the increasing marketization of different LPWAN technologies, such as Sigfox or LoRaWAN. The localization capabilities of these technologies have been tested in practice. The LoRa alliance has released a geolocalization white paper, presenting an overview of the methods tested and used by its members \cite{LoRaWAN_white_paper}. Similarly, the Sigfox company is advertising the capabilities of its localization service~\cite{Sigfox_geolocation}.
 A detailed comparative analysis of the most prominent LPWAN technologies  can be found in~\cite{Sigfox_Dataset} and references therein.

The architecture of these networks is straightforward. Basestations, which are connected to a central server, are deployed statically in urban and rural areas. Depending on the use-case, the low power devices might be statically deployed as, for instance, sensors repeating measurements in fixed locations, or might be mobile. In the latter case, they could be mounted on vehicles so that they report sensor measurements in various locations, or they could be used for asset tracking. The devices transmit messages formatted according to the protocol of the technology used, which is received by the basestations in range. All messages are centrally gathered to a central server. 

Apart from the content of the message which depends on the use case and the task assigned to the mobile devices, several other types of information concerning the transmission are being reported. These types may be: the Received Signal Strength Indicator (RSSI), the Time of Arrival (ToA), the Time Difference of Arrival (TDoA), the Logarithmic Signal over Noise Ratio (LSNR), etc. This information can be utilized by ranging techniques of localization, such a multilateration, to offer position estimates. Ranging techniques have the advantage that they do not require a surveying phase, and can been used directly, assuming knowledge of the basestations' locations. Nevertheless, since they do not inherently contain information related to the particularities of the environment over which they are used, they have a disadvantage, in terms of accuracy, when compared with fingerprinting methods.

Fingerprinting methods rely on datasets that are collected throughout the area of interest. These datasets contain measurements of signal reception values characterizing known locations. These datasets of fingerprints recorded in known locations are used in order to build models which predict the unknown location of new signal receptions. A disadvantage of fingerprinting techniques is that the creation and maintenance of an up-to-date fingerprint database requires considerable effort and cost. One practical way to record such a dataset outdoors is presented in the work of Aernouts et al.~\cite{Sigfox_Dataset}. In that work, a dataset is made publicly available and its collection methodology is presented. A total of 20 cars of the Belgian postal services were equipped with low power devices communicating via Sigfox and LoRaWAN with a central server. At the same time, the location of the car, as estimated by a GPS device, was also reported. In these datasets, the GPS estimates are considered as the spatial ground truth of the location of the message transmission.

Recent works in the field of indoor and outdoor positioning ~\cite{Sigfox_Dataset}, \cite{IndoorLoc_IPIN2017}, \cite{Adler_survey} have underlined the fact that it is indispensable for the publications of the field to favour the reproducibility of the experiments and the comparability of the results.
% As stated in~\cite{IndoorLoc_IPIN2017},`\textit{In the Pattern Recognition and Machine Learning research fields, the common practice is to test the results of each proposal using several well-known datasets.}' 
In this spirit, we utilize a publicly available Sigfox dataset~\cite{Sigfox_Dataset}, and we share with the community the code of the experiments of the current study (DOI:10.5281/zenodo.3228752), and the train/validation/test set split used in said dataset (DOI: 10.5281/zenodo.3228744), to facilitate a consistent comparison of results in future works.

In the current study, we present a detailed examination of the hyperparameter tuning and of the process of selecting the most appropriate preprocessing methodologies for RSSI fingerprinting on an urban Sigfox setting. A systematic examination of the data preprocessing and the hyperparameter tuning steps can optimize the performance of a localization system. Consequently, it may offer an evaluation of the capabilities of the technology used. This work aims to exemplify such a process, and to characterize and report the capabilities of a Sigfox-based localization system on a well defined urban setting.

The rest of this paper is organized as follows.
In Section~\ref{sec:Related}, the work relevant to this subject is discussed. Section~\ref{sec:Dataset} presents in detail the dataset used in this work. In Section~\ref{sec:Preprocessing} the preprocessing steps analysed in this work are presented. After a concise presentation of the experimental setup in Section~\ref{sec:Setup}, an extensive presentation and discussion of the results is developed in  Section~\ref{sec:Results}. Finally, conclusions drawn and ideas for future work are presented in Section~\ref{sec:Conclusions}.
	
\section{Related Work} \label{sec:Related}

Fingerprinting techniques have been common ground for the indoor positioning community for the last two decades~\cite{WiFi_Fingerprint_Overview}. Particularly, RSSI has been the main measurement type that is used ~\cite{WiFi_Fingerprint_Overview}. Wi-Fi and Bluetooth have been some of the most commonly used technologies for indoor fingerprinting~\cite{Survey_Smartphones}. Indoor localization using smartphones has been a main field of application of fingerprinting techniques, in which context fingerprinting is either used as a standalone method or combined with other methodologies in hybrid systems ~\cite{Survey_Smartphones}.

The proliferation of Low Power Wide Area Networks (LPWAN), such as Sigfox and LoRaWAN, has brought a new domain of application of the fingerprinting methods. A recent study~\cite{Aernouts_Comparison} has experimentally verified the intuitive assumption that fingerprinting methods outperform, in terms of accuracy, proximity or ranging positioning methods, in a Sigfox setting. 

Several works have studied localization methods using LPWAN technologies. Plets et al.~\cite{Plets}, have evaluated experimentally RRS and TDoA ranging positioning methods using a LoRaWAN network, reporting median errors of 1250 and 200 meters for RRS and TDoA respectively. Similarly, Podevinj et al.~\cite{Podevijn} present a tracking algorithm which uses a TDoA ranging method for its initial raw estimates that feed the tracking module, reporting a similar median performance (200m) on the raw localization estimates. The two aforementioned works however do not report the mean error of their systems.
Other works~\cite{Choi},\cite{Gotthard}, have focused on rather specific settings over which they evaluate positioning methods. The first work~\cite{Choi} evaluates an RSSI fingerprinting algorithm in the confined area of a parking lot, with four basestations installed in this area, reporting a mean error of 24 meters. It is noteworthy that since knn fingerprinting methods cannot predict a location outside the area of the fingerprints of the training set, a very limited area of fingerprint collection will drastically affect the error statistics. The later work~\cite{Gotthard}, also deals with the subject of car parking, but with a different approach, as in this case the cars were equipped with tags that exchanged messages among them, only later reporting the RSS to a server. In a small-scale test presented, a maximal error of 8 meters is claimed.

In a series of studies~\cite{Sigfox_Dataset},~\cite{Aernouts_Comparison},~\cite{Janssen_Sigfox}, the authors of the IDLab of the University of Antwerp have presented a publicly available Sigfox dataset and have elaborated on the proper configuration of a kNN fingerprinting algorithm.
In the publication in which the dataset used in this study was presented to the research community~\cite{Sigfox_Dataset}, the authors exemplified the usage of the dataset, reporting a 689 meter mean error in this first approach  of utilizing the dataset. In a following publication of the IDLab by Janssen et al.~\cite{Janssen_Sigfox}, the authors present a detailed study of numerous distance metrics and data preprocessing (or data representation) methods, using the same dataset and achieving a mean error of 340 meters. The current study builds on top of the ground these works have set, further examining certain parameters of the preprocessing steps and presenting the impact that an appropriate tuning of these parameters can have in the reduction of the positioning error.

\section{The Dataset Used} \label{sec:Dataset}

%\textcolor{blue}{9/05/2017  1/2 pages}

Aenrouts et al.~\cite{Sigfox_Dataset}  have made publicly available 3 fingerprinting datasets of Low Power Wide Area Networks. Two of these datasets were collected in the urban area of Antwerp, one using Sigfox and another using LoRaWAN. The third dataset was collected in the rural area between the towns of Antwerp and Ghent, using Sigfox. The authors underline their motivation by mentioning in their work that:
`\textit{With  these  datasets,  we  intend  to  provide the  global research  community  with  a benchmark  tool  to  evaluate  fingerprinting algorithms  for LPWAN standards.}'

\begin{figure}[!h]
\centering
\includegraphics[width=0.99\linewidth]{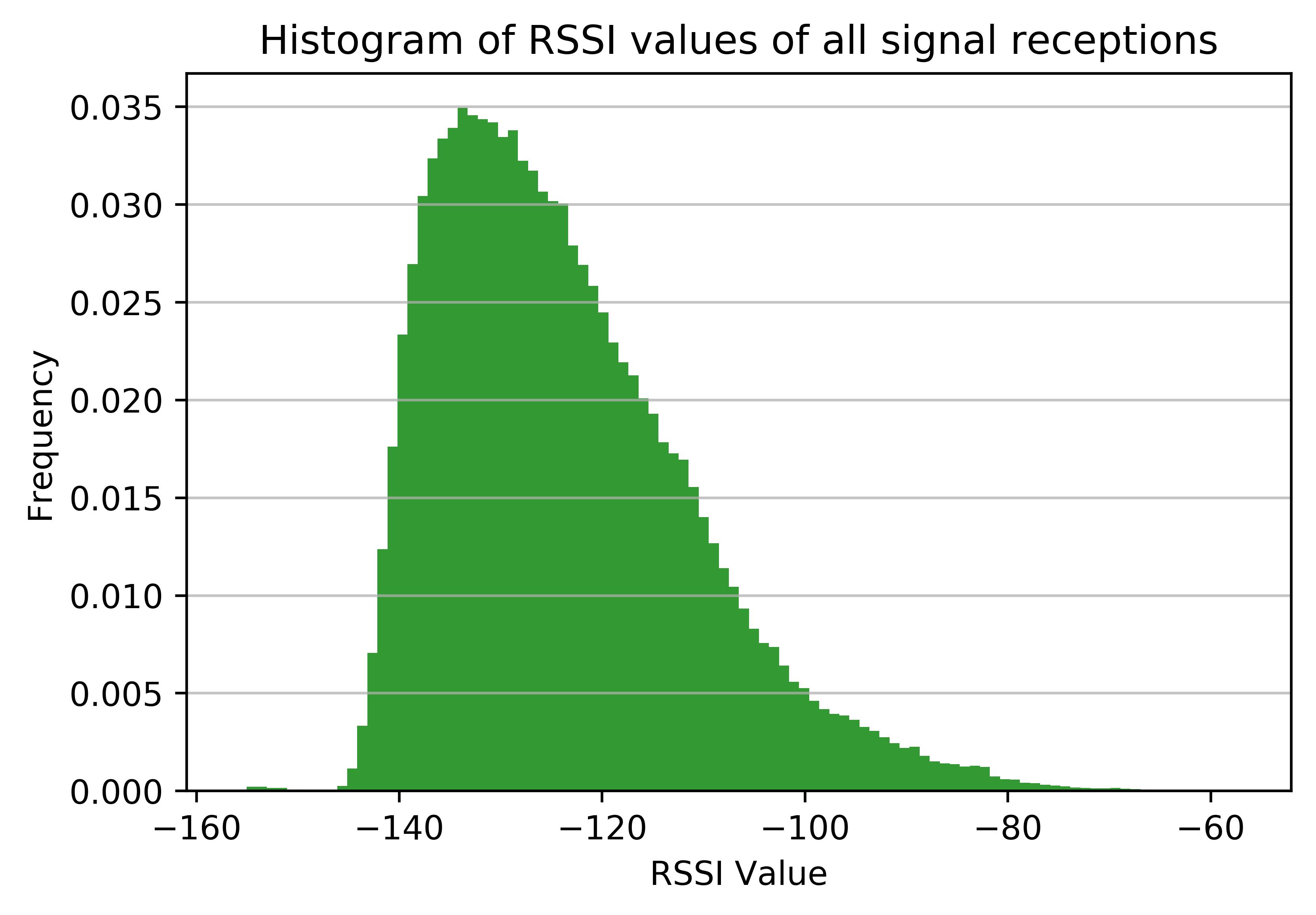}
\caption{The histogram of RSSI values of all signal receptions.
} \label{fig:histogram_rssi_values.png}
\end{figure}

 In this work, we have used the first dataset, containing Sigfox messages in the urban area of Antwerp. The fingerprints are collected in an area of approximately 53 square kilometers, though the majority of them lay in the central area of Antwerp which is approximately half the size of the full area. A total number of 14378 messages are reported in the dataset. Each message contains the following information: the RSSI value of the transmitted signal by each of the 84 base stations, and the spatial ground truth of the signal's transmission location, as estimated by a GPS device, alongside the Sigfox devices. Undeniably, the fact that an estimate which is subject to error is used as ground truth introduces bias. Nevertheless, the error of GPS is at the order of a few tens of meters while the localization accuracy ranges at the order of several hundreds of meters.

In order to obtain a feeling of the distribution of the RSSI values of the dataset, the histogram of all 317126 RSSI values that are present in the dataset is plotted in Figure~\ref{fig:histogram_rssi_values.png}. A big part of the distribution (more that 60\% of the data) is concentrated in the $[-140,-120]$ range, having almost 10\% of the data in the value range above $-100$.
 In cases where a basestation did not receive a message, an out of range RSSI value of $-200$ was inserted, in order not to leave an empty value and also for differentiating this entry from the minimum RSSI values actually received ($-156$). The out of range values of $-200$ were not included in the histogram's creation.

\begin{figure*}[!t]
\centering
\includegraphics[width=0.99\linewidth]{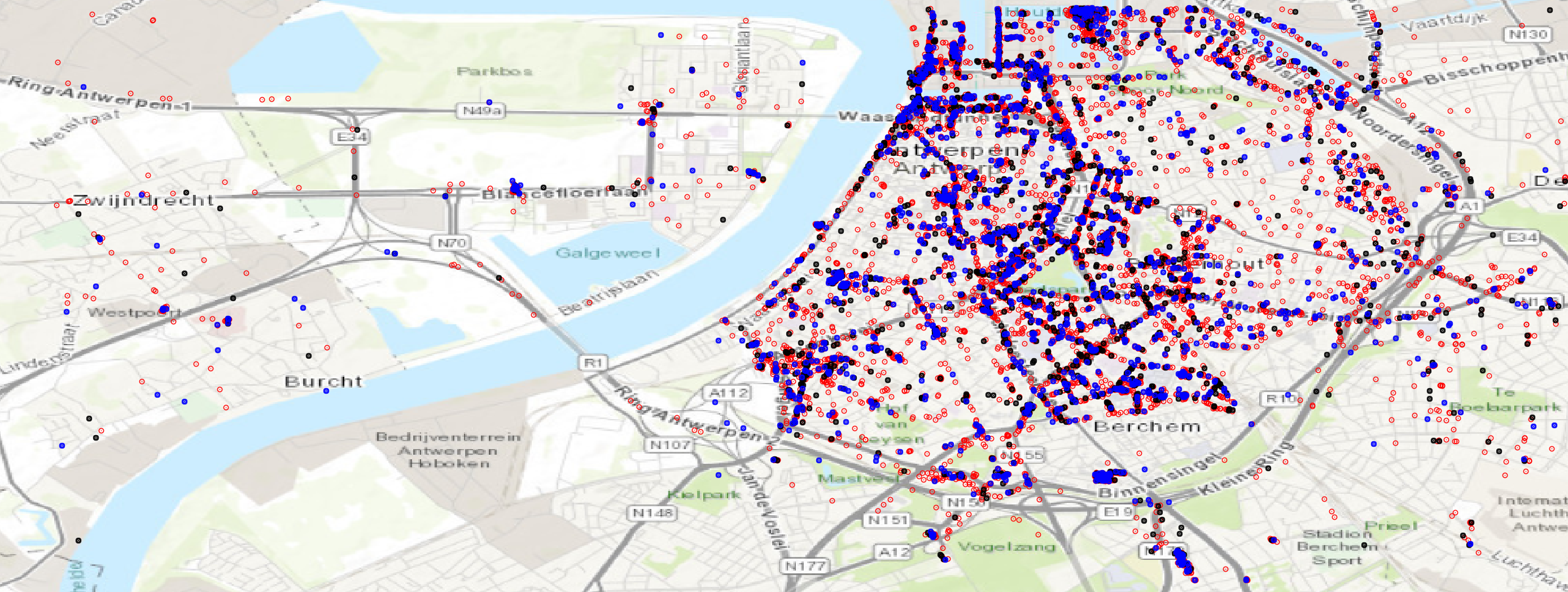}
\caption{The spatial distribution of the data points of the dataset, as reported by the GPS devices used to establish the ground truth. The train, validation and tests sets that were created in this work are depicted in red, black and blue respectively.
} \label{fig:train_val_test_map.png}
\end{figure*}

For the purposes of our study, we have split the dataset into a training, a validation and a test set containing 70\%, 15\%, and 15\% of the sample respectively (10063, 2157 and 2157 entries, in absolute numbers). The spatial distribution of the ground truth locations of the three aforementioned sets can be seen in Figure~\ref{fig:train_val_test_map.png}.

In the spirit of verifiability, reproducibility and comparability of results, it is important not only to report performance metrics over a publicly available dataset, but to also report the specific way the dataset is split into a training, a validation and a test set. In this way, researchers will be able to train models on the same training set, make decisions over the optimal tuning at the same validation set, and most importantly report unbiased performance results on the same test set. The three subsets that are used in the current study are made available to the research community for future reference. For the cases that a different training and validation strategy is desired, or an entirely different dataset is to be used, the full code implementation is available so that the same tests can be reproduced in these different settings.

\section{Data Preprocessing} \label{sec:Preprocessing}  

In a highly cited work, Torres-Sospedra et al.~\cite{TORRES_SOSPEDRA_2015} have presented a systematic study of data preprocessing methods, and of distance and similarity metrics for Wi-Fi fingerprinting in indoor positioning systems. In that work, the authors have shown the great significance of an appropriate data preprocessing step, and have presented four alternative methods of preprocessing the fingerprint data. These four data representation methods are the following:

As defined in~\cite{TORRES_SOSPEDRA_2015}, \textit{positive} values data representation subtracts the minimum RSSI value from all the entries.

\begin{equation}
        Positive_i\\(x)=   \left\{ \begin{array}{ll}
           (RSS_i - min) & \; \parbox{9em}{if basestation $i$ received the message and $RSS_i \geq \tau$}  \\ 
            0 &\text{otherwise} \\
        \end{array} \right.
\end{equation}

%
%           (RSS_i - min) & \textrm{if basestation i received the message and} RSS_i \geq \tau \\ 

where $RSS_i$ is the RSS from the  $i_{th}$ basetation, $min$ is the lowest RSS value minus 1 among all RSS values of the database~\cite{TORRES_SOSPEDRA_2015}. 
Also, $\tau$  is a threshold value, used so that basetations with intensities lower than $\tau$ are considered as not-detected, and the lowest possible value is assigned to them. 

It is worth emphasizing here that, since utilizing in the training set's preprocessing any kind of information coming from the validation or the test set would introduce \textit{information leakage}, the $min$ should only be calculated by the training set data. Calculating the $min$, or normalizing according to the data of the full dataset, would imply a breach of protocol of the train/validation/testing scheme, where the validation and testing sets are supposed to be completely unknown during the preprocessing phase.

%In this work, we slightly alter the definition of \textit{positive} values, presenting  it in a two-step manner. Our definition assumes that there are no empty values in the dataset, meaning that when a signal was not received by a basestation an out-of-range value is used, as it is the case for the aforementioned dataset. The two-step process is:

\begin{itemize}
  \item Replace all values $RSS_I<\tau$ with $\tau$, where  $\tau \geq min$
  \item Define $Positive_i(x)$ as:
\end{itemize}

\begin{equation}
        Positive_i(x)= RSS_I-\tau
        \label{Eq:positive_new}
\end{equation}

The \textit{normalized} data representation, normalizes the \textit{positive} values data representation into the [0,1] range. (To be noted that the RSSI values are negative)

\begin{equation}
        Normalized_i\\(x)=  Positive_i\\(x)/ (-min)
        \label{Eq:Normalized}
\end{equation}

The \textit{exponential} and \textit{powed} representations are the result of the intention to go beyond the linear handling of the RSSI values, since the RSSI values correspond to a logarithmic scale. More particularly, the \textit{exponential} and \textit{powed} representations are defined as follows:

\begin{equation}
        Exponential_i\\(x)= \frac{exp( \frac{Positive_i\\(x)}{\alpha})}{exp( \frac{-min}{\alpha})}
        \label{Eq:Exponential}
\end{equation}

\begin{equation}
       Powed_i\\(x)= \frac{(Positive_i\\(x))^\beta}{(-min)^\beta)}
        \label{Eq:Powed}
\end{equation}

The proposed default values for the parameters $\alpha$ and $\beta $ are  $\alpha=24 $ and $\beta = e $, where $e$ is the mathematical constant. Nevertheless, since these parameters were selected upon testing with Wi-Fi signals in an indoor setting, which has a different range and distribution of RSSI values, it is a good advice to adjust them in accordance with a new setting. This ground setting analysis of data processioning presented in~\cite{TORRES_SOSPEDRA_2015}, has been used as a reference in the work by Janssen et al.~\cite{Janssen_Sigfox}, and is studied in further detail in the current work.
The impact of these preprocessing steps and of the appropriate tuning of their parameters on the achieved positioning accuracy is extensively discussed in Section~\ref{sec:Results}, along with all results of this study.

%\section{Experimental Setup and Results } \label{sec:Setup_and_Results }  

\section{Experimental Setup} \label{sec:Setup}

%\textcolor{blue}{22/05/2019  3? pages}

%\textcolor{red}{Talking about sklearn, our code providing the  train/val/test sets	splits and standards random states. }

In this study, a detailed examination of various prepossessing methods is presented, along with a hyperparameter tuning of the k-nearest neighbours (kNN) method. For the experiments presented in this study the free machine learning library for the Python language, scikit-learn, was used. Particularly, scikit-learn version 0.19.1 and Python 3.5.5 were used. The Haversine formula has been used for measuring distances on the reported experiments. Upon completion of the experiments, the Vincenty formula, which gives the shortest geodesic distance on an ellipsoid modeled earth, was also tested. It is straightforward that the comparisons among the performance of models remain consistent if either of the two formulas is used. The absolute distances measured with the two formulas differ by less than 0.5\%.

The common train/validation/test set division methodology is used on the analyses presented in this work. The train set is used to train the model. In the case of knn, no training process needs to take place per se, thus the training set  practically creates the space of neighbors among which each new fingerprint is compared against. The validation set is used for evaluating different candidate models and checking the appropriateness of the selected hyperparameter values and preprocessing steps. Based on the performance on the validation set, the optimal model configuration is chosen. As both train and validation sets have participated in the configuration of the final model, an unbiased final evaluation of the model's performance needs to take place using a third set, containing previously unseen data: the test set.
Therefore, during all preprocessing, training and tuning steps, the test set is inaccessible, and no information stemming from it is to be used. 

Initially, similarly to the work by Janssen et al.~\cite{Janssen_Sigfox}, we examine various distance metrics (the full list of the Distance Metrics class of scikit-learn) and tune the hyperparameter k for each distance metric. We perform this analysis twice: once using the dataset as it is with the out-of-range $-200$ value unchanged, and then again, replacing it with the experimentally found minimum of actually received RSSI values in the training set minus one ($-157$). Moreover, we examine the performance of a wide range of candidate values of threshold value $\tau$ of Equation~\ref{Eq:positive_new}. Lastly, concerning the parametric preprocessing representations, \textit{exponential} and \textit{powed}, a tuning of their respective parameters ($\alpha$ and $\beta$) is performed. 

%\textcolor{red}{Explain result in validation set, for comparing, and finally test error reported.}

\section{Results} \label{sec:Results}

\subsection{Distance Metrics, Data Preprocessing and $k$} \label{sec:Distance_Metrics}

In this test, we examine various distance metrics and tune the hyperparameter $k$ for each distance metric.
In the experiments of the current study, all four data representations defined in Equations~\ref{Eq:positive_new}-\ref{Eq:Powed} were tested as a preprocessing step. The results of these tests, presented in Tables~\ref{Table:minimum_200} and ~\ref{Table:experimental_minimum}, have verified a fact that was observed in previous works as well (\cite{TORRES_SOSPEDRA_2015}, \cite{Janssen_Sigfox}): the \textit{exponential} and \textit{powed} representations clearly outperform the \textit{positive} and \textit{normalized} data representations, for all distance metrics used. 
For simplicity, in Tables~\ref{Table:minimum_200} and~\ref{Table:experimental_minimum} we only report the results of the two most performant methods, \textit{exponential} and \textit{powed}. It should be underlined that selection of the optimal hypermarameter value $k$ and the localization error statistics reported in Tables~\ref{Table:minimum_200} and ~\ref{Table:experimental_minimum} are calculated on the validation set. Moreover, in the context of this test, the default values of the parameters ($\alpha=24 $ and $\beta = e $) of the \textit{exponential} and \textit{powed} representations are used.

Concerning the distance metrics, apart from those reported in Tables~\ref{Table:minimum_200} and~\ref{Table:experimental_minimum}, another family of distance metrics available in scikit-learn was evaluated,  but proven to be entirely unsuitable. That family of metrics such as the Jaccard, Matching, Dice or Kulsinski distance, are intended for boolean-valued vector spaces, setting as True any non-zero entry. Consequently, those representations utilize a binary type of information stating if each basestation has received the message or not. Overall, the results show that, in accordance with the previous works, the Bray-Curtis metric (equivalent to the S{\o}rensen metric mentioned in previous works \cite{TORRES_SOSPEDRA_2015}, \cite{Janssen_Sigfox} offers the best results in terms of accuracy.

Comparing the results of Tables~\ref{Table:minimum_200} and~\ref{Table:experimental_minimum}, the following conclusions can be drown. In Table~\ref{Table:minimum_200}, \textit{exponential} representation is more performant than \textit{powed}, for all distance metrics except for the Canberra distance. On the contrary, in Table~\ref{Table:experimental_minimum} it is \textit{powed} representation that offers better results, for all but one distance metrics. In Table~\ref{Table:minimum_200} the best performance is achieved by the Bray-Curtis metric and the \textit{exponential} data representation, with a mean and median error of 344 and 148 meters respectively. For the results of Table~\ref{Table:experimental_minimum}, the best performance is achieved by the Bray-Curtis metric and the \textit{powed} data representation, with 319 and 123 meters of mean and median error respectively. The corresponding   mean and median values of error on the test set are, 301 and 109 respectively.

\begin{table}[h]
\caption{Localization Error Analysis on the Validation Set, Reporting the Optimal k Value of all Distance Metrics and Two Prepossessing Strategies, with the Default Replacement of the Missing Values with the $\tau=-200$ Value}\label{Table:minimum_200}
\centering\setlength{\extrarowheight}{2pt}
\centering
\begin{tabular}{|*{7}{c|}}
\hline
 \multirowcell{2}{DistanceMetric} & \multicolumn{3}{c|}{Exponential RSS} & \multicolumn{3}{c|}{Powed RSS} \\
 \cline{2-7}
& \makecell{\textbf{k}} 
& \makecell{\textbf{mean}} 
& \makecell{\textbf{median}}
& \makecell{\textbf{k}} 
& \makecell{\textbf{mean}} 
& \makecell{\textbf{median}}\\ \hline

\makecell{euclidean}
& \makecell{9}  & \makecell{355} & \makecell{155} 
&  \makecell{8} & \makecell{387} & \makecell{182} \\

\makecell{manhattan}
& \makecell{5}  & \makecell{360} & \makecell{145} 
&  \makecell{6} & \makecell{389} & \makecell{183} \\
 
\makecell{chebyshev}
& \makecell{7}  & \makecell{428} & \makecell{222} 
&  \makecell{3} & \makecell{506} & \makecell{285} \\ 

%\makecell{minkowski}
%& \makecell{8}  & \makecell{355} & \makecell{153} 
%&  \makecell{8} & \makecell{387} & \makecell{182} \\ 

\makecell{hamming}
& \makecell{10}  & \makecell{1100} & \makecell{950} 
&  \makecell{7} & \makecell{1058} & \makecell{906} \\

\makecell{canberra}
& \makecell{6}  & \makecell{516} & \makecell{316} 
&  \makecell{11} & \makecell{572} & \makecell{368} \\
 
\makecell{braycurtis}
& \makecell{8}  & \makecell{\textbf{344}} & \makecell{\textbf{148}} 
&  \makecell{6} & \makecell{\textbf{364}} & \makecell{\textbf{168}} \\ 

%\makecell{jaccard}
%& \makecell{16}  & \makecell{1616} & \makecell{1601} 
%&  \makecell{15} & \makecell{887} & \makecell{707} \\
%
%
%\makecell{matching}
%& \makecell{16}  & \makecell{1616} & \makecell{1601} 
%&  \makecell{14} & \makecell{893} & \makecell{707} \\ 
%
%
%\makecell{dice}
%& \makecell{16}  & \makecell{1616} & \makecell{1601} 
%&  \makecell{15} & \makecell{887} & \makecell{707} \\
% 
%\makecell{kulsinski}
%& \makecell{16}  & \makecell{1616} & \makecell{1601} 
%&  \makecell{8} & \makecell{1055} & \makecell{864} \\ 
%
%\makecell{rogerstanimoto}
%& \makecell{16}  & \makecell{1616} & \makecell{1601} 
%&  \makecell{14} & \makecell{893} & \makecell{706} \\
%
%\makecell{russellrao}
%& \makecell{16}  & \makecell{1616} & \makecell{1601} 
%&  \makecell{16} & \makecell{1646} & \makecell{1423} \\
% 
%\makecell{sokalmichener}
%& \makecell{16}  & \makecell{1616} & \makecell{1601} 
%&  \makecell{14} & \makecell{893} & \makecell{706} \\ 
%
%\makecell{sokalsneath}
%& \makecell{16}  & \makecell{1616} & \makecell{1664} 
%&  \makecell{15} & \makecell{887} & \makecell{707} \\ 
\hline

\end{tabular}
\end{table}

\begin{table}[h]
\caption{Localization Error Analysis on the Validation Set, Reporting the Optimal k Value of All Distance Metrics and Two Prepossessing Strategies, Having Adjusted the Missing Values to the Experimental Minimum RSSI Value, Minus 1 ($\tau=-157$) }\label{Table:experimental_minimum}
\centering\setlength{\extrarowheight}{2pt}
\centering
\begin{tabular}{|*{7}{c|}}
\hline
 \multirowcell{2}{DistanceMetric} & \multicolumn{3}{c|}{Exponential RSS} & \multicolumn{3}{c|}{Powed RSS} \\
 \cline{2-7}
& \makecell{\textbf{k}} 
& \makecell{\textbf{mean}} 
& \makecell{\textbf{median}}
& \makecell{\textbf{k}} 
& \makecell{\textbf{mean}} 
& \makecell{\textbf{median}}\\ \hline

\makecell{euclidean}
& \makecell{8}  & \makecell{345} & \makecell{148} 
&  \makecell{8} & \makecell{343} & \makecell{141} \\

\makecell{manhattan}
& \makecell{7}  & \makecell{348} & \makecell{138} 
&  \makecell{6} & \makecell{343} & \makecell{136} \\
 
\makecell{chebyshev}
& \makecell{4}  & \makecell{409} & \makecell{191} 
&  \makecell{4} & \makecell{399} & \makecell{175} \\ 

%\makecell{minkowski}
%& \makecell{8}  & \makecell{345} & \makecell{148} 
%&  \makecell{7} & \makecell{344} & \makecell{140} \\ 

\makecell{hamming}
& \makecell{10}  & \makecell{1100} & \makecell{950} 
&  \makecell{7} & \makecell{1065} & \makecell{912} \\

\makecell{canberra}
& \makecell{6}  & \makecell{420} & \makecell{205} 
&  \makecell{8} & \makecell{470} & \makecell{275} \\
 
\makecell{braycurtis}
& \makecell{5}  & \makecell{\textbf{341}} & \makecell{\textbf{126}} 
&  \makecell{6} & \makecell{\textbf{319}} & \makecell{\textbf{123}} \\ 

%\makecell{jaccard}
%& \makecell{16}  & \makecell{1616} & \makecell{1601} 
%&  \makecell{16} & \makecell{886} & \makecell{711} \\
%
%
%\makecell{matching}
%& \makecell{16}  & \makecell{1616} & \makecell{1664} 
%&  \makecell{16} & \makecell{899} & \makecell{716} \\ 
%
%
%\makecell{dice}
%& \makecell{16}  & \makecell{1616} & \makecell{1664} 
%&  \makecell{16} & \makecell{886} & \makecell{711} \\
% 
%\makecell{kulsinski}
%& \makecell{16}  & \makecell{1616} & \makecell{1664} 
%&  \makecell{6} & \makecell{1057} & \makecell{882} \\ 
%
%\makecell{rogerstanimoto}
%& \makecell{16}  & \makecell{1616} & \makecell{1664} 
%&  \makecell{15} & \makecell{899} & \makecell{724} \\
%
%\makecell{russellrao}
%& \makecell{16}  & \makecell{1616} & \makecell{1664} 
%&  \makecell{16} & \makecell{1667} & \makecell{1457} \\
% 
%\makecell{sokalmichener}
%& \makecell{16}  & \makecell{1616} & \makecell{1664} 
%&  \makecell{15} & \makecell{899} & \makecell{724} \\ 
%
%\makecell{sokalsneath}
%& \makecell{16}  & \makecell{1616} & \makecell{1664} 
%&  \makecell{16} & \makecell{886} & \makecell{711} \\ 
\hline

\end{tabular}
\end{table}

 While the results obtained by the \textit{positive} and \textit{normalized} representation were not reported in Tables~\ref{Table:minimum_200} and ~\ref{Table:experimental_minimum}, it is worth to briefly discuss the performance they achieve. The Bray-Curtis distance metric remains the best performing one for both data representations. Since the \textit{normalized} representation is just a rescaled version of the \textit{positive} data representation, both methods achieve identical results in this setting. When the dataset is used as is, meaning with $\tau=-200$, the mean error on the validation set is 552 meters, while for $\tau=-157$, the error is 400 meters. A first observation is that an appropriate transformation of the RSS values may change entirely the level of accuracy, leading from an initial error of 552 meters in the linear handling of the RSS values, to a 344 meters error. Furthermore, adjusting the $\tau$ value, and therefore, the way out-of-range values are set, may further improve the performance, thus the 319 meters mean error observed in Table~\ref{Table:experimental_minimum}.

In an attempt of an interpretation of this difference in performance among the two different threshold values used, the following arguments are presented. It appears that the selection of an out-of-range value may drastically affect the performance of the positioning method. In the studied dataset, the $-200$ out-of-range value is quite distinct to the experimental minimum RSSI found in the dataset ($-156$). This artificial gap between these two values may assign a significant importance to the distinction between a very distant gateway receiving the signal and a non-receiving one. On the other hand, bridging this gap may treat these two cases as more similar, and favor the distinction of RSSI values among closer detected basestations, improving the distance measurement between fingerprints, and consequently the efficient selection of the closest neighbours.

\subsection{Threshold Value $\tau$} \label{sec:Threshold_Value}

In this test, the impact of the value of the $\tau$ threshold on the localization performance is examined. The best configuration found so far is used ($k=6$, with a  \textit{powed} data representation), for examining all $\tau$ values in the range $[-200,-130]$. 
Setting $\tau$ with a value higher than the experimental minimum of $-156$, will replace all values in the range $[-156,\tau]$ with $\tau$. Out-of-range values that are lower that the experimental minimum, are simply set to $\tau$.

\begin{figure}[!h]
\centering
\includegraphics[width=0.99\linewidth]{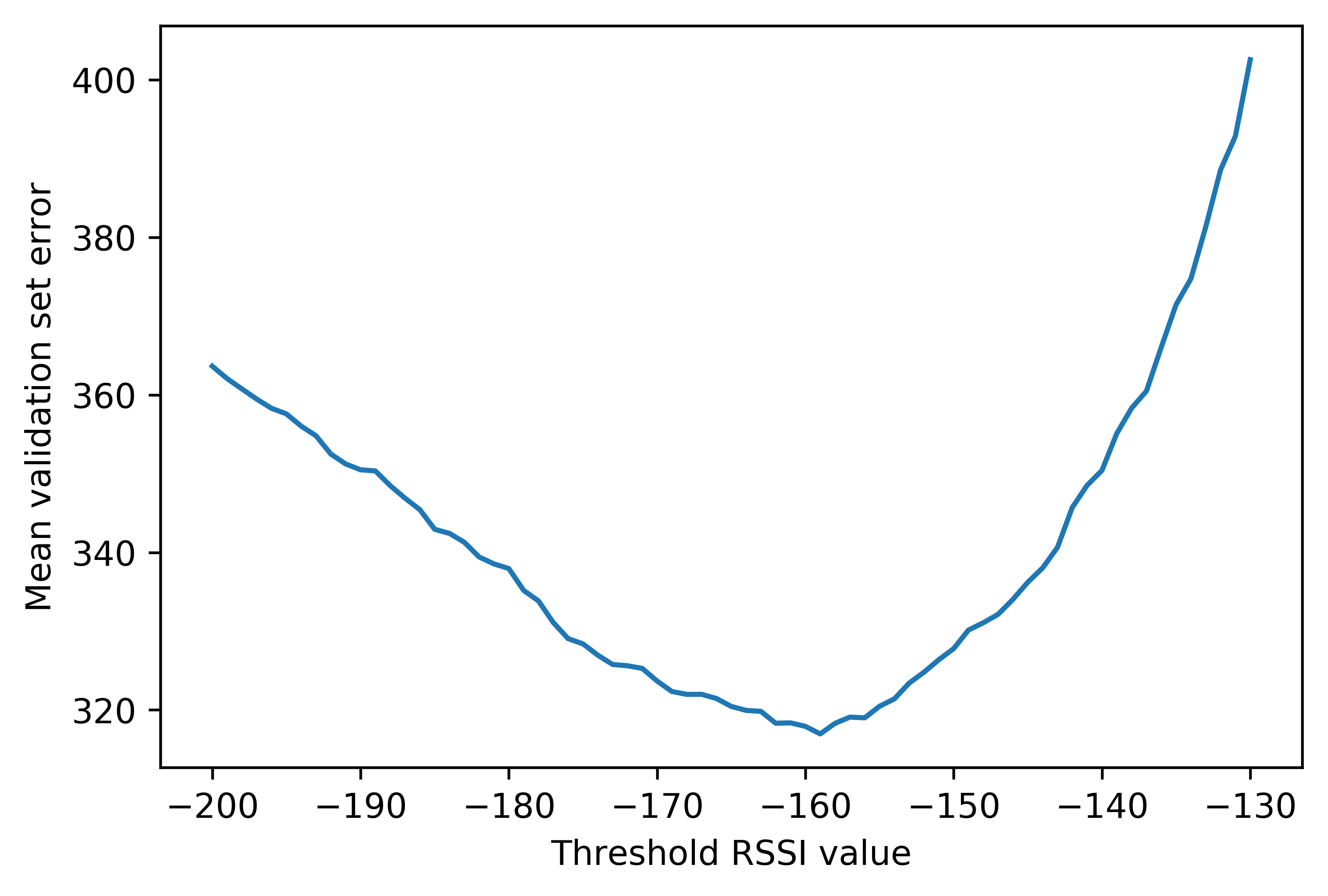}
\caption{The mean error on the validation set for different values of the threshold $\tau$ of Equation~\ref{Eq:positive_new}. 
} \label{fig:Error_for_threshold_RSSI_values.png}
\end{figure}

The results of this analysis are reported in Figure~\ref{fig:Error_for_threshold_RSSI_values.png}. It is evident that the best performance comes from values around the experimental minimum. In particular, the lowest mean error in the validation set was 317 meters, and it was given by $\tau=-159$. The corresponding performance in the test set was 298 meters mean error and 109 meters median error. The optimal value $\tau=-159$ is just below the experimental minimum of received RSSI values, which suggests that putting the out-of-range value just below the experimental minimum is the best option, according to this test.

\subsection{Parameters a and b of the Exponential and Eowed Data Representations} \label{sec:a_and_b}

As presented in Section~\ref{sec:Preprocessing}, the \textit{exponential} and \textit{powed} data representations rely on a parameter ($\alpha $ and $\beta $ respectively). The study that introduced these two representations~\cite{TORRES_SOSPEDRA_2015} recommends default values for these parameters, upon experimentation with data coming from indoor WiFi measurements. The range and distribution of those values are different than those of the dataset used in this work. Therefore, an appropriate adjustment of the values of parameters $\alpha$ and $\beta$ would adapt these data transformations to optimally fit this different setting.
 
\subsubsection{Parameter $\alpha$ of the Exponential Data Representation} \label{sec:a}

\begin{figure}[!h]
\centering
\includegraphics[width=0.99\linewidth]{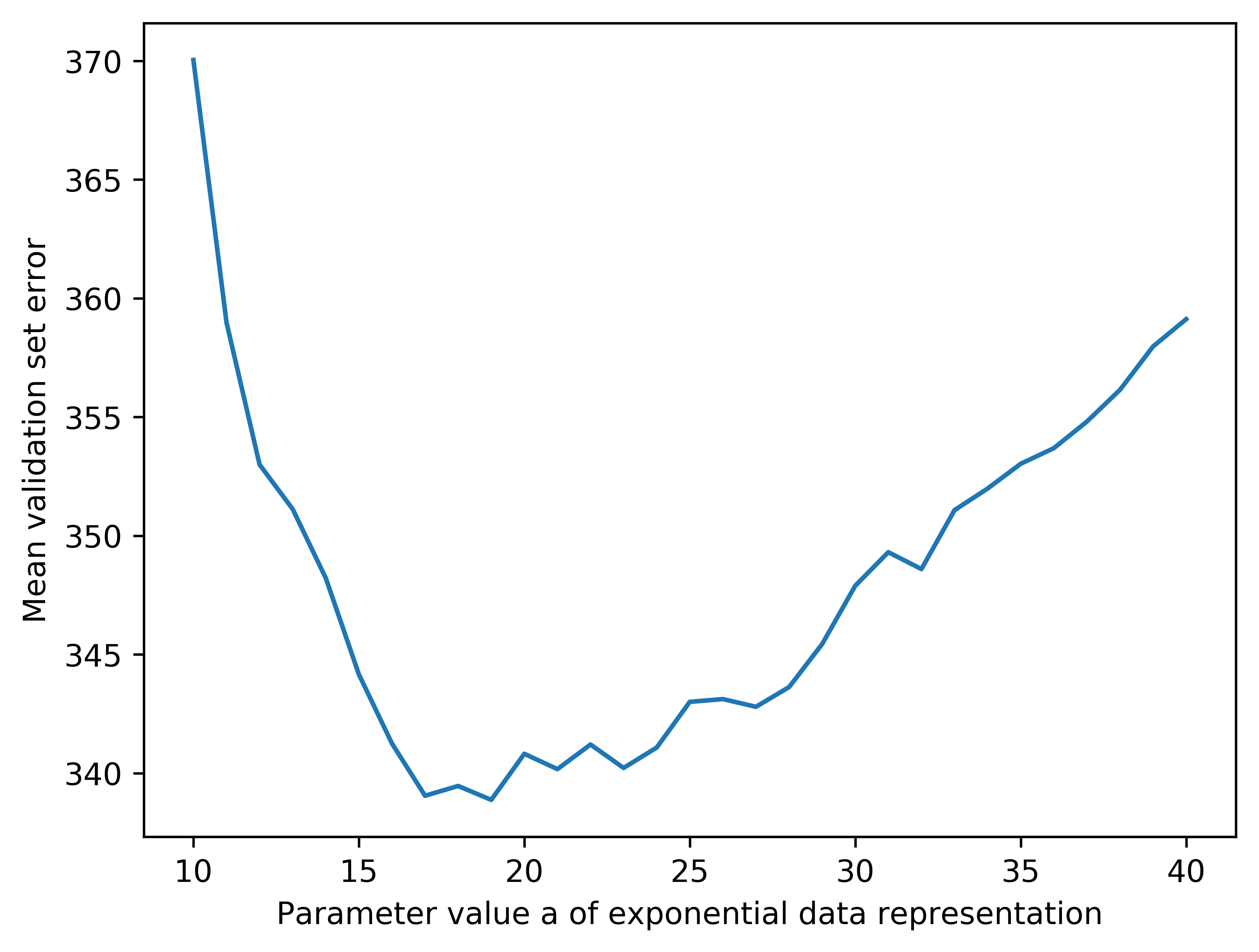}
\caption{The mean error on the validation set for different values of the parameter $\alpha$ of the \textit{exponential} data representation.
} \label{fig:alpha.png}
\end{figure}

Regarding the parameter $\alpha$ of the \textit{exponential} data representation, a range of candidate values has been tested. It is reminded that the default value of $\alpha$ is 24. Integer values in the range [10,40] have been evaluated in the best configuration found so far, concerning the \textit{exponential} data representation.  Thus, the Bray-Curtis distance has been used with $k=5$ and $\tau=-157$. The results are presented in the plot of Figure~\ref{fig:alpha.png}.

The value $\alpha=19$ provides the lowest mean validation error of 339 meters. The corresponding test set performance is characterized by a mean error of 318 meters and a median of 117 meters. In the plot of Figure~\ref{fig:alpha.png} it can be observed that there is a significant difference in performance comparing to the default value ($\alpha=24$).

\begin{figure}[h]
\centering
\includegraphics[width=0.99\linewidth]{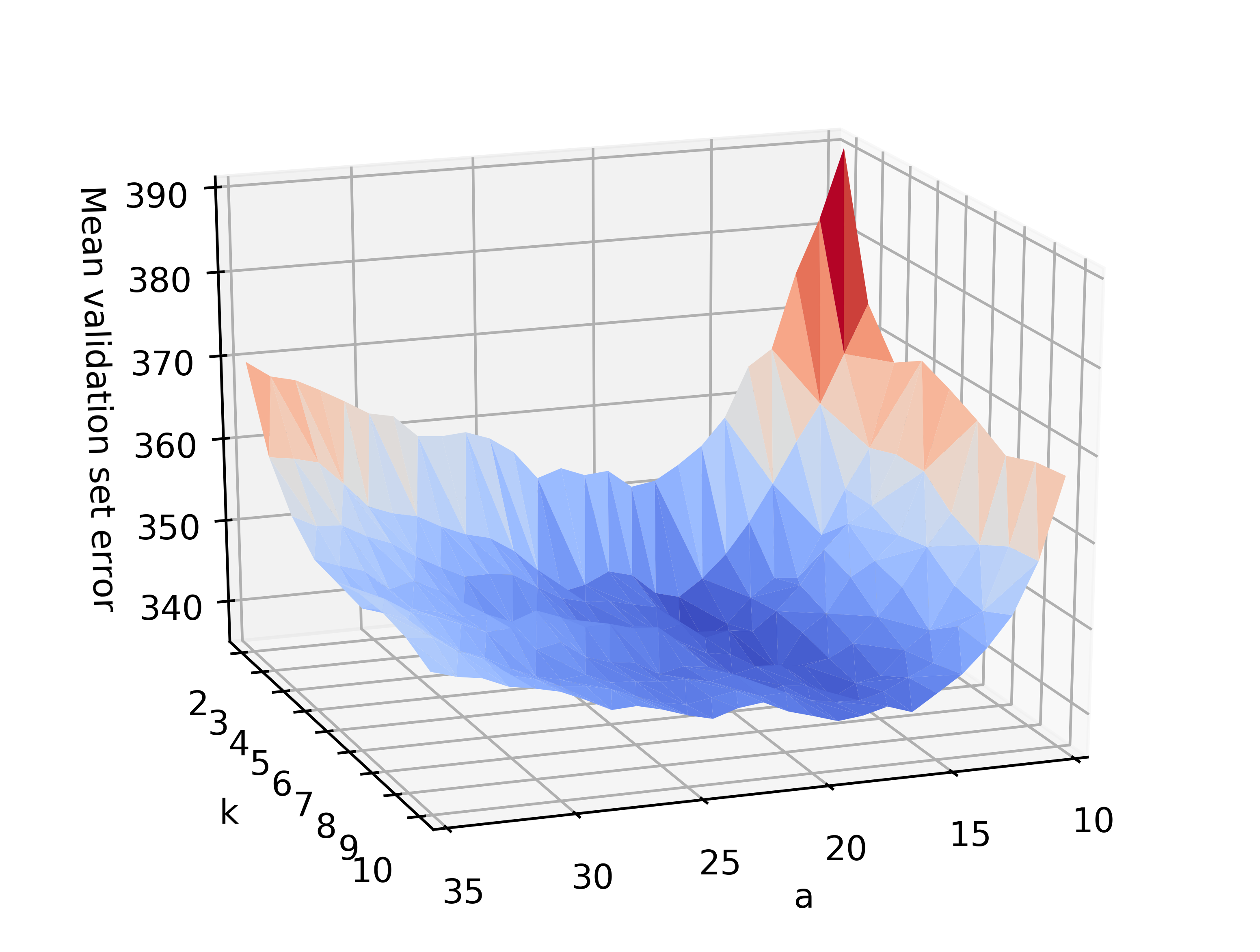}
\caption{The mean error on the validation set for different values of the parameter $\alpha$ of the \textit{exponential} data representation and of the $k$ hyperparameter of kNN.
} \label{fig:alpha_k.png}
\end{figure}

 In Figure~\ref{fig:alpha_k.png}, the results of a similar examination are depicted. This time, the test spans the space of candidate values of both the parameter $\alpha$ of the \textit{exponential} data representation and of the $k$ hyperparameter of kNN. The minimum mean validation error is provided by $\alpha=18$ and $k=4$, and is equal to 336 meters. The corresponding performance in the test set shows a mean error of 322 meters and a median error of 110 meters. There are several tuples of values of $\alpha$ and $k$ in the proximity of the ones reported above that provide very similar results.

\subsubsection{Parameter $\beta$ of the Exponential Data Representation} \label{sec:a}

\begin{figure}[!h]
\centering
\includegraphics[width=0.99\linewidth]{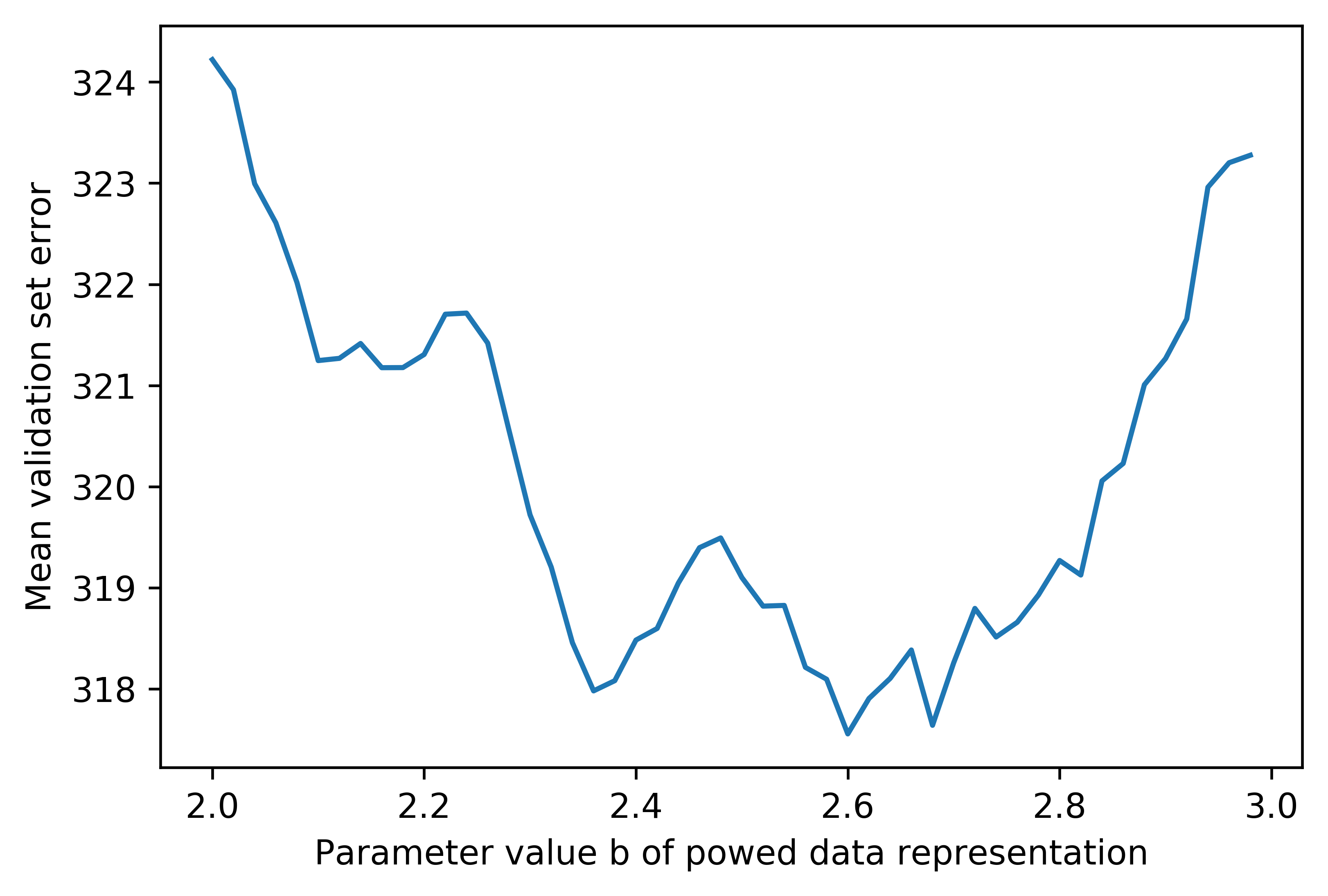}
\caption{The mean error on the validation set for different values of the parameter $\beta$ of the \textit{powed} data representation.
} \label{fig:beta.png}
\end{figure}

An analysis similar to the previous one is performed for  the parameter $\beta$ of the \textit{powed} data representation. The range [2,3] has been spanned with a granularity of 0.02. The default value of $\beta$ is the constant $e$, which is equal to 2.718. The best configuration found concerning the \textit{powed} data representation has been used. Thus, the Bray-Curtis distance has been used with $k=6$. The results are presented in the plot of Figure~\ref{fig:beta.png}.

The value $\beta=2.6$ provides the lowest mean validation error of 318 meters. The mean error on the test set is 298 meters and the median 108 meters. In this case that parameter $\beta$ is studied, the difference of performance of the best  $\beta$ value compared to the one of the default value, is less significant than in the relevant analysis of $\alpha$ that preceded. Indicatively, it is noted that the default value of  $\beta$ gave a mean error of 319 meters.

\begin{figure}[!h]
\centering
\includegraphics[width=0.99\linewidth]{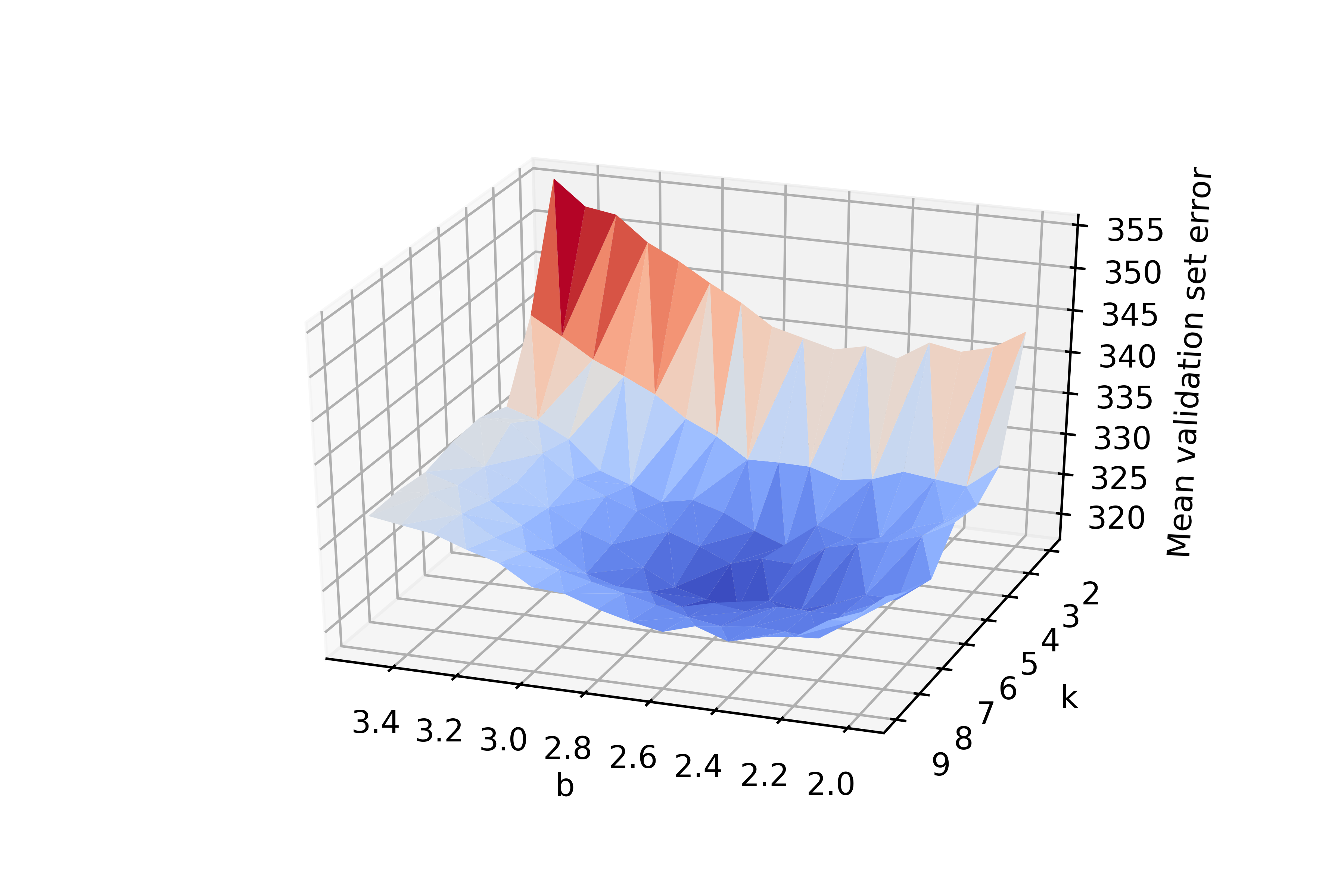}
\caption{The mean error on the validation set for different values of the parameter $\beta$ of the \textit{powed} data representation and of the $k$ hyperparameter of kNN. 
} \label{fig:beta_k.png}
\end{figure}

 In Figure~\ref{fig:beta_k.png}, the results for combinations of candidates values of both the parameter $\beta$ of the \textit{powed} data representation and of the $k$ hyperparameter of kNN are reported. The minimum mean validation error is provided by $\beta=2.6$ and $k=6$, same as in the previous analysis of Figure~\ref{fig:beta.png}. The convex form surface of Figure~\ref{fig:beta_k.png} reveals that the lowest values of the mean error appear in a close proximity , since there are several tuples of values $\alpha$ and $k$ in the proximity of the optimal ones reported above that provide very similar results.

% 
%\begin{figure}[h]
%\centering
%\includegraphics[width=0.99\linewidth]{alpha.png}
%\caption{The mean error on the validation for different values of the threshold RSSI value. 
%} \label{fig:histogram_rssi_values.png}
%\end{figure}

\section{Conclusions and Future Work}
\label{sec:Conclusions}

In this work, we presented a detailed study of the process of selecting the most appropriate preprocessing methodologies and performing hyperparameter tuning for RSSI fingerprinting in an urban Sigfox setting. We have discussed the ways of appropriately preprocessing of the RSSI data, so as to improve the  accuracy of the studied fingerprinting localization method. Moreover, identifying the limits of the achievable accuracy of a Sigfox-based positioning system deployed in an urban area has a been a main motivation of this work.

 %Utilizing a publicly available dataset, an emphasis has been given to the reproducibility of the tests and the comparability of the results, and therefore the code and the train/validation/test set split used in this work are publicly available.

The examination of the results of this study may offer several take-away messages. A linear handling of the RSSI values, with the \textit{positive} and \textit{normalized} data representations, achieved an above 500-meter mean error on the validation set. Introducing non-linear transformations, proposed by the relevant literature as more appropriate for the handling of RSSI values which correspond to a logarithmic scale, reduced the error to the level of 344 meters. To the best of our knowledge, this study is the first go beyond these steps and further optimize the handling of out-of-range values and the tuning of the parameters $\alpha $ and $\beta $ of the preprocessing data transformation, so as to match the particularities of this outdoors setting, which does not utilize a Wi-Fi system as the original paper~\cite{TORRES_SOSPEDRA_2015}, but another technology, namely Sigfox. The best performing setting proposed in this work achieves a mean error of 317 meters on the validation set. This setting achieves a 298-meter error on the test set, with the corresponding median error being 109 meters. 

A comparison with the performance of previous works would not be strictly consistent, even if the same dataset has been used, since the validation and test sets are not the same. Nevertheless, we could discuss the order of magnitude of the error achieved. In the initial work where the used dataset became available~\cite{Sigfox_Dataset}, the linear handling of the RSSI value resulted in a high mean error of 689 meters. A more recent work~\cite{Janssen_Sigfox}, has utilized the data preprocessing methods proposed by Torres-Sospedra et al.~\cite{TORRES_SOSPEDRA_2015}, reporting a mean validation error of 340. The results of Table~\ref{Table:minimum_200} of subsection~\ref{sec:Distance_Metrics}, report a similar mean error of 344 meters, for the same best data representation found in ~\cite{Janssen_Sigfox}, the \textit{exponential}. 

The aditional analysis of the threshold value $\tau$, and the appropriate  adjustment of the preprocessing parameters $\alpha$ and $\beta$ proposed by this work, further improves the localization accuracy. Moreover, it is noteworthy that the appropriate adjustment of $\tau$, sets a different preproseccing method as the best performing one. While in Table~\ref{Table:minimum_200} of subsection ~\ref{sec:Distance_Metrics}, as well as in the work of Janssen et al.~\cite{Janssen_Sigfox}, the \textit{exponential} data representation is suggested by the obtained results, the appropriate adjustment of $\tau$ sets the \textit{powed} data representation as preferable. It is these preprocessing adjustments that give a mean error of 317 meters on the validation set and 298 meters in the test set, with 121 and 109 meters for the corresponding median errors.

The improvements in accuracy obtained by firstly using and secondly tuning these preprocessing steps indicate the significance of these steps. An appropriate preprocessing of the data may have huge impact on the performance of the positioning system using them, and thus it is recommended as an indispensable step of the model selection process

A driving force of motivation of the authors has been the intention to encourage the sharing of material among the positioning community, which can facilitate consistent comparisons, and accelerate the advancement of the field. Being thankful for the public offering of the dataset to the community by Aernouts et al.~\cite{Sigfox_Dataset}, we proceed in sharing the dataset's split into train/test/validation sets, used in this work, as well as the code used for the tests. 

As localization with LPWANs is a rather recent field of study, there aren't many publicly available datasets so far. Moreover, the size of the dataset plays a crucial role, when machine learning approaches are used for localization. A dense spatial sampling of the area of interest can positively affect the localization performance. Additionally, other types of measurements, apart from the RSSI value, such as ToA, TDoA or LSNR values, would be of great interest to be studied. The intention of the authors is to work in the direction of creating and sharing such datasets. We also invite the community to embrace this effort that can accelerate the improvement of the field.

Lastly, an interesting future direction for studies such as the current one may be the evaluation of the computational complexity of the machine learning methods and the distance metrics that are tested. In the current work, the focus has been on the performance of the positioning system in terms of accuracy. Nevertheless, in practical settings factors such as the computational complexity may play a crucial role in the selection of the model used.

\section*{Acknowledgements}
\label{sec:Acknowledgements} 
This work was funded by the Commission for Technology and Innovation CTI, of the Swiss federal government, in the frame of the OrbiLoc project (CTI 27908.1 PFES-ES).

%
% \clearpage 
\balance

\bibliographystyle{IEEEtran}
\bibliography{bibliography/bibliography}

\end{document}